\documentclass[twocolumn,showpacs,nofootinbib,prd,aps,amsmath,amssymb,superscriptaddress]{revtex4-1}
\usepackage{graphicx,epsfig}
\usepackage{bm}
\usepackage{color}
\newcommand{\ba}{\begin{eqnarray}}
\newcommand{\ea}{\end{eqnarray}}

\newcommand{\ACal}{{\cal{A}}}

\newcommand{\DD}{{\cal {D}}}

\newcommand{\bbq}{\begin{quote}}
\newcommand{\eeq}{\end{quote}}

\newcommand{\tbb}{t_{\textrm{\tiny{bb}}}}

\newcommand{\RR}{{}^{(3)}{\cal{R}}}

\newcommand{\FF}{{\cal{F}}}
\newcommand{\GG}{{\cal{G}}}
\newcommand{\JJ}{{\cal{J}}}
\newcommand{\VV}{{\cal{V}}}
\newcommand{\HH}{{\cal{H}}}
\newcommand{\KK}{{\cal{K}}}
\newcommand{\PP}{{\cal{P}}}

\newcommand{\Da}{\Delta^{(A)}}

\newcommand{\Drho}{\Delta^{(\rho)}}
\newcommand{\DKK}{\Delta^{(\KK)}}
\newcommand{\dKK}{\delta^{(\KK)}}
\newcommand{\drho}{\delta^{(\rho)}}
\newcommand{\Drhoas}{\Delta^{(\rho)}_{(\textrm{\tiny{as}})}}

\newcommand{\Drholin}{\Delta^{(\rho)}_1}

\newcommand{\DDa}{{\textrm{\bf{D}}}^{(A)}}
\newcommand{\DDrho}{{\textrm{\bf{D}}}^{(\rho)}}
\newcommand{\DDKK}{{\textrm{\bf{D}}}^{(\KK)}}

\newcommand{\DDTh}{{\textrm{\bf{D}}}^{(\Theta)}}
\newcommand{\DDalin}{{\textrm{\bf{D}}}_1^{(A)}}

\newcommand{\DDKKlin}{{\textrm{\bf{D}}}_1^{(\KK)}}

\newcommand{\DDThlin}{{\textrm{\bf{D}}}_1^{(\Theta)}}

\newcommand{\ddKK}{{\textrm{\bf{d}}}^{(\KK)}}
\newcommand{\ddTh}{{\textrm{\bf{d}}}^{(\Theta)}}

\newcommand{\ddkk}{{\textrm{\bf{d}}}^{(\kappa)}}

\newcommand{\DDaas}{{\textrm{\bf{D}}}^{(A)}_{(\textrm{\tiny{as}})}}

\newcommand{\DDKKas}{{\textrm{\bf{D}}}^{(\KK)}_{(\textrm{\tiny{as}})}}
\newcommand{\DDThas}{{\textrm{\bf{D}}}^{(\Theta)}_{(\textrm{\tiny{as}})}}

\newcommand{\DrhoLS}{\Drho_{1\textrm{\tiny{LS}}}}
\newcommand{\DDKKLS}{\DDKK_{1\textrm{\tiny{LS}}}}

\newcommand{\eb}{{\hbox{\bf{e}}}}

\newcommand{\bW}{{\rm{\bf W}}}

\newcommand{\tls}{t_{\textrm{\tiny{LS}}}}
\newcommand{\Hls}{\bar H_{\textrm{\tiny{LS}}}}

\newcommand{\Ommls}{\bar \Omega^m_{\textrm{\tiny{LS}}}}
\newcommand{\OmLls}{\bar \Omega^\Lambda_{\textrm{\tiny{LS}}}}
\newcommand{\dd}{{\rm{d}}}

\newcommand{\Cg}{C_{\tiny{\textrm{(+)}}}}
\newcommand{\Cd}{C_{\tiny{\textrm{(--)}}}}
\newcommand{\Jg}{\JJ_{\tiny{\textrm{(+)}}}}
\newcommand{\Jd}{\JJ_{\tiny{\textrm{(--)}}}}

\begin{document}
\title[Non--spherical Szekeres models in the language of Cosmological Perturbations]{{Non-Spherical} Szekeres models in the language of Cosmological Perturbations} 
\author{Roberto A. Sussman} 
\affiliation{Instituto de Ciencias Nucleares, Universidad Nacional Aut\'onoma de M\'exico,
A. P. 70--543, 04510 M\'exico D. F., M\'exico,}
\email{sussman@nucleares.unam.mx}
\author{Juan Carlos Hidalgo}
\affiliation{Instituto de Ciencias F\'\i sicas, Universidad Nacional Aut\'onoma de M\'exico, 62210 Cuernavaca, Morelos, M\'exico,}
\email{hidalgo@fis.unam.mx}
\author{Ismael Delgado Gaspar}
\affiliation{Instituto de Investigaci\'on en Ciencias B\'asicas y Aplicadas, Universidad Aut\'onoma del Estado de Morelos, Av. Universidad 1002, 62210 Cuernavaca, Morelos, M\'exico.}
\author{Gabriel Germ\'an}
\affiliation{Instituto de Ciencias F\'\i sicas, Universidad Nacional Aut\'onoma de M\'exico, 62210 Cuernavaca, Morelos, M\'exico,}
\date{\today}
\begin{abstract} 
We study the differences and equivalences between the non--perturbative description of the evolution of cosmic structure furnished by the Szekeres dust models (a non--spherical exact solution of Einstein's equations) and the dynamics of Cosmological Perturbation Theory (CPT) for dust sources in a $\Lambda$CDM background. We show how the dynamics of Szekeres models can be described by evolution equations given in terms of ``exact fluctuations'' that identically reduce (at all orders) to evolution equations of CPT in the comoving isochronous gauge. We explicitly show how Szekeres linearised exact fluctuations are specific (deterministic) realisations of standard linear perturbations of CPT given as random fields but, as opposed to the latter perturbations, they can be evolved exactly into the full non--linear regime. We prove two important results: (i) the conservation of the curvature perturbation (at all scales) also holds for the appropriate linear approximation of the exact Szekeres fluctuations in a $\Lambda$CDM background, and (ii) the different collapse morphologies of Szekeres models yields, at nonlinear order, different functional forms for the growth factor that follows from the study of redshift space distortions. The metric based potentials used in linear CPT are computed in terms of the parameters of the linearised Szekeres models, thus allowing us to relate our results to linear CPT results in other gauges. We believe that these results provide a solid starting stage to examine the role of non--perturbative General Relativity  in current cosmological research.                                                   
\end{abstract}
\pacs{98.80.-k, 04.20.-q, 95.36.+x, 95.35.+d}
\maketitle
\section{Introduction}
Gauge invariant perturbations on a Friedman--Lema\^\i tre--Robertson--Walker (FLRW) background, generically examined within the framework of Cosmological Perturbation Theory (CPT), constitute an important theoretical tool in cosmological research (see pioneering work in \cite{CPTh} and more recent comprehensive reviews in \cite{CPT1,CPT2,CPT3}). Linear CPT is specially adequate to study cosmic sources whenever near homogeneous conditions can be justified: the early Universe and scales comparable to the Hubble radius for late cosmic times. Since late time structure formation at deep sub--horizon scales becomes highly nonlinear and (at least locally) nonÑrelativistic, it is usually studied by means of non--linear and non--perturbative Newtonian gravity, either through analytic models (spherical \cite{newt1,newt2,newt3,barrow-silk} and elliptic collapse \cite{barrow-silk,newt4,newt5})  or by sophisticated numerical N-body simulations \cite{numsim1,numsim2}. By considering higher order perturbations (Newtonian and relativistic) CPT has been so far extended to study the mildly nonlinear regime \cite{CPT21,CPT22,CPT23,CPT24,CPT25,CPT26}, leaving the description of fully nonlinear effects in large scale structure formation to Newtonian non--perturbative methods, though the need for considering relativistic corrections is still an open question (an extensive review is \cite{CPT2summary}). 

A proper understanding of the evolution of the density and the peculiar velocities in cosmic inhomogeneities is essential to distinguish between specific cosmological models.  While several dark energy or modified gravity background FLRW models reasonably fit  late time observational data, it is the clustering properties of matter that allows us to distinguish between these models \cite{Lahav-Rees,Bueno-Bellido,Percival-Song}.  The growth of structure is  characterised by a growth factor $f$ function computed in linear CPT, which is  an observable derived from the anisotropy of the power spectrum in redshift space in the non--linear regime \cite{Scoccimarro}. Yet, it is precisely at this non--linear regime that departures from Newtonian evolution arise, which suggests considering the evolution of inhomogeneities at non--linear level by introducing corrections from  General Relativity (GR) theory.

Non--perturbative and fully non--linear GR theory is not a particularly favoured theoretical tool in cosmological research, since (given its high nonlinear complexity) any minimally realistic non--perturbative and relativistic modelling of structure formation necessarily requires numerical 3--dimensional codes to solve Einstein's equations in a cosmological context, whether as a continuum model or as GR numerical simulations. This impressively difficult task is still in its early stages of development \cite{GRN1,GRN2,GRN3,GRN4,GRN5,GRN6,GRN7}. However, all numerical and perturbative work (whether Newtonian or relativistic) still requires simple analytic ``toy models'' that provide useful practical hints and comparative qualitative results. From a fully relativistic and non--perturbative approach, these toy models emerge from appropriate exact solutions of Einstein's equations. Since cold dark matter (CDM) at cosmic scales can be adequately described by a dust source and a $\Lambda$ term can always be added to the dynamics as a phenomenological description of dark energy, the most useful  exact GR solutions applicable to cosmology are the well known spherically symmetric dust Lema\^\i tre--Tolman--Bondi (LTB) models \cite{ltbold} and their non--spherical generalisation furnished by Szekeres models \cite{Sz751,Sz752} (see comprehensive reviews in \cite{kras1,kras2,BKHC2009,BCK2011,EMM} for both classes of solutions).

The {usage} of LTB and Szekeres models in cosmological applications can (and should) also be undertaken within the prevailing $\Lambda$CDM paradigm or Concordance Model. It is important to emphasise this point, since employing these {exact solutions} is often associated with the recent past attempt to  challenge this paradigm by means of non--perturbative large scale (Gpc sized) density void configurations mostly constructed with LTB models \cite{LTBV1,LTBV2} and, to a lesser degree, with simplified Szekeres models (see reviews in \cite{kras1,kras2,BKHC2009,BCK2011,EMM}). Once LTB void models were ruled out \cite{LTBV3,LTBV4,LTBV5}, thus re--affirming the validity of the Concordance Model, interest on cosmological application of exact GR solutions decreased. However, the claim that non--perturbative GR is redundant for cosmological research (within or without the $\Lambda$CDM paradigm) is an open issue \cite{GRE1,GRE2,GRE3,GRE4,GRE5}. As long as the  cosmological implementation of numerical GR remains under development, it should be appropriate to continue applying LTB and Szekeres models as theoretical tools to complement perturbative and numerical cosmological research. 

Because of their spherical symmetry LTB models only allow us to describe the evolution of a single CDM structure (an overdensity or a density void) imbedded in a suitable FLRW background (see numerical examples in \cite{BKHC2009}). Evidently, Szekeres models introduce more dynamical degrees of freedom, as can be appreciated in the above cited reviews and in theoretical studies \cite{GW,barrow1,barrow2,theo1,theo2,theo3,theo4,bolsus,sussbol,buckley,theo5, multi}, as well as in the extensive literature on their application to structure formation and fitting of cosmological observations \cite{SFO1,SFO2,SFO3,SFO4,SFO5,SFO6,SFO7,SFO8,SFO9,SFO10,SFO11,SFO12,SFO13, SFO14,SFO15,SFO16,coarse}. Practically all of these articles (see exceptions below) only consider the simplest type of structure formation scenario allowed by the models: a 2--structure dipolar configuration comprised by an overdensity evolving next to a density void  (see numerical examples in \cite{BKHC2009}). The possibility of modelling less restrictive structures was suggested already in earlier work \cite{bolsus,buckley,SFO1,SFO2}, but it only became recently implemented \cite{multi,coarse} by using the full extension of the dynamical freedom of the models for the description of elaborated networks consisting of an arbitrary number of CDM structures (overdensities and voids).

Since Szekeres models (specially those examined in \cite{multi,coarse}) provide the less idealised exact GR solution in a cosmological context, they are specially {suitable} to examine the connection between non--perturbative GR and CPT based perturbations at different orders, as well as with non--perturbative Newtonian models. In the present article we explore this connection through a detailed rigorous comparison between the dynamical equations of Szekeres models and CPT for dust sources in the comoving isochronous gauge, thus extending and continuing a previously published \cite{perts} similar comparison between LTB models and CPT. This represents the first stage in an effort to compare all these theoretical tools in terms of their structure modelling predictions and their usage in fitting observations and addressing open theoretical issues.

The contents of this article are summarised as follows. Section 2 introduces the metric of Szekeres models and their particular cases (spherical and axial symmetry) in spherical coordinates. In sections 3 and 4 we show how the dynamics of the models can be completely determined by suitable covariant variables defined in \cite{sussbol} and used in \cite{multi,coarse}: the q--scalars associated with the density, the Hubble scalar and spatial curvature, their corresponding FLRW background variables and exact fluctuations that convey the inhomogeneity of the models. The evolution equations for these variables (as we show) can be adequately related to the CPT evolution equations as in the LTB case \cite{perts}. Employing these variables is crucial, as the standard metric based description of Szekeres model (as used in all previous work on these models save for \cite{sussbol,multi,coarse}) is not useful for relating to CPT dynamics.  We define in section 5 a linear regime in Szekeres models by suitable first order expansions of the exact fluctuations as done with LTB models in \cite{perts}. In particular, we show that this linear regime is compatible with large deviations from spherical symmetry that allow for a description of networks of multiple evolving structures (overdensities and voids \cite{multi,coarse}). The linearised Szekeres evolution equations and their solutions are examined in section 6. 

In section 7 we examine the equivalence between the Szekeres and CPT evolution equations for dust sources in the isochronous CPT gauge. We follow the methodology of \cite{perts} but with more depth and generality: while we focus primarily on first order linear equations, we show that the Szekeres evolution equations are equivalent to CPT equations at all approximation orders. We prove that the first order Szekeres spatial curvature fluctuation is conserved in time for all scales, just as its equivalent linear spatial curvature perturbation from CPT. 

In section 8 we describe the ``pancake'' and spherical collapse morphologies allowed by Szekeres models, then we compute the Szekeres analogue of the growth suppression factor $f$ used in CPT to address the study of redshift space distortion \cite{Percival-Song,Scoccimarro,RSD1,RSD2}. We show that this analogue fully coincides with its linear equivalent from CPT and that is only sensitive to collapse morphologies in a non--linear regime (either second order CPT or exact). In section 9 we obtain the Szekeres equivalent CPT metric potentials (in the isochronous gauge) by linearising the Szekeres line element in terms of its deviation from the FLRW metric. This allows us to relate our dynamical equations with CPT quantities in any gauge. 

In section 10 we present concluding remarks and a discussion of the main results listed above.  We also present a conceptual comparison between the CPT density perturbation, given as a random field, and its Szekeres equivalent: the linearised Szekeres density fluctuation, given as a linearised expression that follows from a deterministic exact solution. We argue that the linear CPT perturbation is far more general, but is only valid in the mildly non--linear regime, whereas its equivalent linearised Szekeres fluctuation can be evolved into a fully non--perturbative exact expression valid throughout the models evolution.

Finally, we provide two appendices: \ref{AppA} discusses the integration of the Friedman--like quadrature that determines the functional form of the metric variables (the scale factors), while \ref{AppB} derives the linearised form of these scale factors.  

\section{Szekeres models and their sub--cases.}

We consider  quasi--spherical Szekeres models of class I in terms of ``stereographic'' spherical coordinates \cite{kras2,multi,coarse} 
\footnote{All further mention of ``Szekeres models'' will refer only to quasi--spherical models of class I (see \cite{kras2} for a broad discussion on their classification). The standard diagonal metric form these models and the transformation relating it to (\ref{szmetric}) is given in Appendix A of \cite{multi}. For constant time slices with spherical or wormhole topology \cite{kras2} the metric (\ref{szmetric}) must be modified as explained in Appendix D of \cite{sussbol}.}
\ba 
\dd s^2&=& -\dd t^2 +h_{ij}\,\dd x^i\,\dd x^j,\qquad i,j=r,\theta,\phi,\nonumber\\
h_{ij}&=&h_{\mu\nu}\delta^\mu_i\delta^\nu_j=g_{\mu\nu}\delta^\mu_i\delta^\nu_j=a^2\gamma_{ij}, \quad a=a(t,r),
\label{szmetric}
\ea
where $h_{\mu\nu}=g_{\mu\nu}+u_\mu u_\nu,\,\,u^\mu=\delta^\mu_t$ and the nonzero components of $\gamma_{ij}$ are:
\ba
\gamma_{rr} &=& \frac{(\Gamma-\bW)^2}{1-\KK_{qi}r^2}+(\PP+\bW_{,\theta})^2+U^2\PP_{,\phi}^2,\label{szmetric1}\\
 \gamma_{r\theta} &=& - r\,(\PP+\bW_{,\theta}),\quad \gamma_{r\phi} =  r\,\sin\theta\,U\,\PP_{,\phi},\label{szmetric2}\\
 \quad\gamma_{\theta\theta} &=& r^2,\quad \gamma_{\phi\phi}=r^2\,\sin^2\theta,\label{szmetric3}
\ea
with the functions $\Gamma,\,U,\,\PP$ and $\bW$ given by
\ba 
\Gamma &=& 1+\frac{ra'}{a},\qquad U=1-\cos\theta,\label{aux1}\\
\PP &=& X\cos\phi+Y\sin\phi,\quad \bW=-\PP\sin\theta-Z\cos\theta,\label{aux2}
\ea
where the free parameters $X,\,Y,\,Z,\,\KK_{qi}$ depend only on $r$ (see the interpretation of $\KK_{qi}$ in (\ref{qscals})), and $'$ denotes the radial derivative.  The function $\bW$ has the mathematical structure of a dipole whose orientation is governed by the choice of the three dipole parameters $X,\,Y,\,Z$ (see comprehensive discussion in \cite{multi}). Different particular cases follow by specialising these parameters:
\begin{description}
\item[Spherical Symmetry: LTB models.] 
We can define for each Szekeres model the {\bf ``LTB seed model''} as the generic LTB model that follows as its spherical limit: $X=Y=Z=0\,\,\Rightarrow\,\,\bW=0$, whose metric is the following specialisation of (\ref{szmetric}) 
\begin{equation} \gamma_{rr} = \frac{\Gamma^2}{1-\KK_{qi}r^2},\quad \gamma_{\theta\theta} = r^2,\quad \gamma_{\phi\phi}=r^2\,\sin^2\theta.\end{equation}
Notice that any Szekeres model can be constructed from an arbitrary LTB seed model simply by introducing suitable nonzero dipole parameters $X,\,Y,\,Z$. Evidently, Szekeres models always ``inherit'' the properties of their LTB seed models, a feature that is very useful to study their properties.
\item [Axial Symmetry.] Another important particular case is furnished by axially symmetric models: $X=Y=\PP=0,\,\,Z\ne 0$, leading to the specialised metric
\ba
\gamma_{rr} &=& \frac{(\Gamma-\bW)^2}{1-\KK_{qi}r^2}+\bW_{,\theta}^2,\quad \gamma_{r\theta} = - r \bW_{,\theta},\notag \\
 \gamma_{\theta\theta} &=& r^2,\quad \gamma_{\phi\phi}=r^2\,\sin^2\theta,\label{axial}
 \ea
where $\bW=-Z\,\cos\theta$ is independent of $\phi$. The asymptotic spherical limit follows if $Z\to 0$ as $r\to \infty$.
\end{description}

\section{Dynamics through quasi--local scalars}

Practically all work considering cosmological applications of Szekeres models \cite{kras1,kras2,BKHC2009,BCK2011,EMM} examine the dynamics of the models in terms of the metric variables determined   from a Friedman quadrature that follows directly from Einstein's field equations (see \ref{AppA} and \ref{AppB}). This approach is not useful to relate the models to CPT. Instead, we determine the dynamical equations (including the metric functions) through the first order system of evolution equations derived in \cite{sussbol} and used in \cite{multi,coarse}
\ba  \dot\rho_q &=& -3 \rho_q\,\HH_q,\label{FFq1}\\
 \dot \Theta_q &=& -\frac{\Theta_q^2}{3}-4\pi\rho_q+8\pi\Lambda, \label{FFq2}\\
 \dot\Delta^{(\rho)} &=& -(1+\Drho)\,\DDTh\label{FFq3}\\
 \dot {\textrm{\bf{D}}}^{(\Theta)} &=&  \left(-\frac{2}{3}\Theta_q+\DDTh\right)\DDTh-4\pi\rho_q\Drho,\label{FFq4}\\
 \dot a &=& a\,\frac{\Theta_q}{3}, \label{FFq5}\\
 \dot\GG &=& \GG\,\DDTh,\qquad \GG=\frac{\Gamma-\bW}{1-\bW}, \label{FFq6}\ea
with $\dot{} = \partial / \partial t$, and with quantities subject to the algebraic constraints:
\ba
\left(\frac{\Theta_q}{3}\right)^2 &=&\frac{8\pi}{3}\left[\rho_q+\Lambda\right]-\KK_q,\label{constraints1}\\
\frac{3}{2}\DDKK &=& {4\pi}\rho_q\Drho-\frac{\Theta_q}{3}\DDTh.\label{constraints2}\ea
Here the quasi--local (q--scalars) $A_q$ and their exact fluctuations $\DDa$ \cite{sussbol} are given for each scalar $A=\rho,\,\Theta,\,\KK,$ (density, Hubble expansion and spatial curvature) by
\ba 
&&A_q = \frac{\int_\DD{A\,\Xi\,\dd \VV_p}}{\int_\DD{\FF\,\dd \VV_p}}, \label{Aqdef}\\
 &&\DDa = A-A_q=\frac{r\,A'_q}{3(\Gamma-\bW)},\, \label{DaDrho1}\\ 
 &&\Drho = \frac{\DDrho}{\rho_q}=\frac{\rho-\rho_q}{\rho_q},\label{DaDrho2}\\
\text{with}&&\qquad\dd\VV_p = \sqrt{\hbox{det}(g_{ij})}\,\dd^3x \notag \\
&&\qquad\qquad= \frac{a^3\,r^2\,(\Gamma-\bW)\,\sin\theta}{\Xi}\,\dd r\dd\theta\dd\phi,\label{Aqint}
 \ea
 and 
where $\Xi=\sqrt{1-\KK_{qi}r^2}$. This last integral in (\ref{Aqdef}) is evaluated in an arbitrary time slice (constant $t$) in a spherical comoving domain $\DD$ bounded by an arbitrary fixed $r>0$. This leads to the scaling laws
\footnote{The lower bound of the integrals (\ref{Aqdef}) is the locus $r=0$, analogous to the symmetry centre of spherical models \cite{bolsus}. While Szekeres models are not spherically symmetric, the surfaces of constant $r$ are non--concentric 2--spheres \cite{kras2,multi}. Notice that $A_q=A_q(t,r)$ even if the scalars $A$ depend on the four coordinates $(t,r,\theta,\phi)$ \cite{theo3,bolsus,sussbol}. Their relation with the average integrals is discussed in \cite{bolsus,sussbol}.}
\ba \rho_q &=& \frac{\rho_{qi}}{a^3},\quad \KK_q=\frac{\KK_{qi}}{a^2},\quad \frac{\Theta_q}{3}=\frac{\dot a}{a},\label{qscals}\\
1+\Drho &=& \frac{1+\Drho_i}{\GG},\quad \frac{2}{3}+\DKK =\frac{\frac{2}{3}+\DKK_i}{\GG},\label{qperts}\ea      
where $\GG$ is defined in \eqref{FFq6} and the subindex ${}_{i}$ denotes evaluation at an arbitrary time slice $t=t_{i}$.  

The first order evolution equations for the fluctuations $\Drho$ and $\DDTh$ can be combined into a single second order equation
\begin{equation}\ddot\Delta^{(\rho)}-\frac{2\left[\dot\Delta^{(\rho)}\right]^2}{1+\Drho}+\frac{2}{3}\Theta_q\,\dot\Delta^{(\rho)}-4\pi\rho_q\,\Drho\,\left(1+\Drho\right)=0,\label{q2ndorder}\end{equation}
which resembles an exact (non--linear) generalisation of the equation of linear dust perturbations in the synchronous gauge \cite{perts}. 

The initial conditions to integrate the system (\ref{FFq1})--(\ref{constraints2}) are specified at $t=t_i$ and consist of the cosmological constant $\Lambda$ plus the following five free functions that depend only on $r$: 
\ba\hbox{The ``radial'' functions:}\,\,\rho_{qi},\,\KK_{qi}.\label{initconds1}
\\
\hbox{The ``dipole'' functions:}\,\,X,\,Y,\,Z.\label{initconds}\ea
where the radial functions are common to the LTB seed model and the dipole functions  govern the deviation from spherical symmetry. We obtain the initial values of $\Theta_q,\,\Drho,\,\DDTh,\,\DDKK$ by solving the constraints  (\ref{constraints1})--(\ref{constraints2}) at $t=t_i$, the radial coordinate is chosen so that $a_{i}=\Gamma_{i}=1$, while the Big Bang time $\tbb$ and its gradient $\tbb'$ can be obtained from $\rho_{qi},\,\KK_{qi}$ and their gradients (see \cite{sussbol,multi,coarse} and \ref{AppA}, \ref{AppB}). 

\section{Dynamics through contrast perturbations.}\label{contrast}

The fluctuations $\DDa$ and $\Da$ in (\ref{DaDrho1})--\eqref{DaDrho2} compare the scalars $A=\rho,\,\HH,\,\KK$ with their associated q-scalars $A_q=\rho_q,\,\HH_q,\,\KK_q$ at the same values of $r$ for all $t$. As a consequence, the density fluctuation $\Drho$ is different from the notion of a ``density contrast''. However, assuming the existence of an asymptotic FLRW background (see conditions for this in \ref{AppB}), we can define exact fluctuations that yield the notion of a contrast by comparing the scalars $A$ at very point with their background values $\bar A = \bar\rho,\,\bar\Theta,\,\KK$
\footnote{Notice that the covariant background scalars $\bar A$ can be understood as asymptotic limits as $r\to\infty$ of the q--scalars $A_q$. In other words: the $\bar A$ are averages for an asymptotic averaging domain covering the whole time slice (see comprehensive discussion in \cite{perts}).}
. This leads to the following ``asymptotic'' fluctuations
\begin{equation} \DDaas= A-\bar A,\qquad \Drhoas  = \frac{\rho-\bar\rho}{\bar\rho},\end{equation}
with the $\bar A$ given by (FLRW quantities will be henceforth denoted by a overbar)
\begin{equation} \bar\rho=\frac{\bar\rho_i}{\bar a^3},\quad\frac{\bar\Theta}{3}=\frac{\dot{\bar a}}{\bar a}=\bar H,\quad\bar\KK =\frac{\bar\KK_i}{\bar a^2},\qquad \bar a(t)=\lim_{r\to\infty}a,\label{flrwscals}\end{equation}
The corresponding evolution equations and constraints are
\begin{align}
\dot{\bar \rho} &= -\bar\rho\,\bar \Theta,\label{as1}\\
\dot{\bar\Theta} &= -\frac{\bar\Theta^2}{3}-4\pi\bar\rho
+8\pi\Lambda,\label{as2}\\
\dot\Drhoas &= -\left[1+\Drhoas\right]\,\DDThas,\label{as3}\\
\dot {\textrm{\bf{D}}}^{(\Theta)}_{(\textrm{\tiny{as}})} &= -\left[2\bar\Theta-\frac{4}{3}\Theta_q+\DDThas\right]\,\DDThas-\frac{2}{3}\left(\Theta_q-\bar\Theta\right)^2-4\pi\bar\rho\,\Drhoas,\label{as4}
\end{align}
\ba
\left(\frac{\bar\Theta}{3}\right)^2 &=&\frac{8\pi}{3}\left[\,\bar\rho+\Lambda\,\right]-\bar\KK,\label{c1as}\\
\frac{3}{2}\DDKKas &=& {4\pi}\bar\rho\Drhoas-\frac{\Theta_q}{3}\DDThas + \frac{1}{6}\left(\Theta_q-\bar\Theta\right)^2,\label{c2as}\ea
while the equivalent to the second order equation (\ref{q2ndorder}) is
\begin{align}
\ddot\Drhoas-\frac{\left[\dot\Drhoas\right]^2}{1+\Drhoas}+
\left[2\bar\Theta-\frac{4}{3}\Theta_q\right]\,\dot\Drhoas-&\notag \\
\left[4\pi\bar\rho\,\Drhoas -2(\Theta_q-\bar\Theta)^2\right]\,\left(1+\Drhoas\right)&=0,\label{c2ndorder}
\end{align}
which not only ``resembles'' but strictly provides the exact (non--linear) generalisation of the evolution equation of linear dust density perturbation in the synchronous gauge \cite{perts}.

The system (\ref{as1})--(\ref{c2as}) is not self--contained, it thus needs to be supplemented by (\ref{FFq1})--(\ref{FFq2}). However, the contrast fluctuations in (\ref{as1})--(\ref{c2as}) are related to the quasi--local fluctuations defined in the previous section by   
\begin{align}
\Drhoas -\Drho =& \left(\frac{\rho_q}{\bar\rho}-1\right)\left(1+\Drho\right)\nonumber\\
 =& \left[\frac{\rho_{qi}}{\bar\rho_i}\,\frac{\bar a^3}{a^3}-1\right]\,\frac{1+\Drho_i}{\GG},\label{pertseq1}\\
 \frac{1}{3}\left(\DDThas-\DDTh\right) =& \frac{1}{3}\left(\Theta_q-\bar\Theta\right) = \frac{\dot a}{a}-\frac{\dot{\bar a}}{\bar a}.\label{pertseq2}
 \end{align}
Therefore, for all purposes it is more practical to solve first the background equations (\ref{as1})--(\ref{as2}) and then use the solutions of (\ref{FFq1})--(\ref{constraints2}) and (\ref{FFq5})--(\ref{FFq6}) to compute the density contrast and Hubble scalar fluctuation from the relations \eqref{pertseq1}--\eqref{pertseq2}.

\section{The Szekeres linear regime.}


The FLRW limit of Szekeres models can be defined rigorously and in a coordinate independent manner \cite{kras2,multi} by the vanishing of the shear and electric Weyl tensors in the background: $\sigma^a_b=E^a_b=0$,  while their fluctuations are both expressible as    
\ba
 E^\mu_\nu=\Psi_2\,\eb^\mu_\nu,\quad&&\text{with}\quad \Psi_2 =-\frac{4\pi}{3}\,\DDrho,\label{sigE1}\\
\sigma^\mu_\nu=\Sigma\,\eb^\mu_\nu,\quad &&\text{with}\quad\Sigma= -\frac{\DDTh}{3},\label{sigE2}
\ea
where $\eb^\mu_\nu=\hbox{diag}[0,-2,1,1]$ is a unique traceless tensor basis satisfying $ \dot\eb^\mu_\nu=0$ and the fluctuations\\ $\DDrho=\rho_q\Drho$ and $\DDTh$ are
\begin{align}
\DDrho =&\frac{\rho_{qi}}{a^3}\,\frac{1-\Gamma+\drho_i}{\Gamma-\bW},\label{DDrho}\\
\DDTh =& \frac{4\pi\rho_{qi}(1-\Gamma+\drho_i)-a\left[\KK_{qi}(1-\Gamma)+\frac32\ddKK_i\right]}{ a^3\,\,(\Theta_q/3)\,(\Gamma-\bW)},\label{DDTh}
\end{align}
where we used (\ref{constraints2}) and (\ref{qscals})--(\ref{qperts}) and $\drho_i,\,\ddKK_i$ are the initial fluctuations of the LT seed model given by
\begin{align}
 \drho_i = \Drho_i|_{_{\bW=0}}=\frac{r\rho'_{qi}}{3\rho_{qi}},\qquad \ddKK_i=\DDKK_i|_{_{\bW=0}}=\frac{r\KK'_{qi}}{3},\label{LTBperts}
 \end{align}%
where we used the fact that $a_i=\Gamma_i=1$. Proceeding as in LTB models in \cite{perts}, we define a linear regime for Szekeres models (understood as functional parameter ``closeness'' to a FLRW background, see \ref{AppB}) by demanding that a positive dimensionless number $\epsilon\ll 1$ exists such that
\begin{align}
 \hbox{all of}&\quad |\Drho|,\,\,|\DDrho|,\,\,|\DDTh|,\,\,|\DDKK|\, \sim O(\epsilon)\notag \\
& \Rightarrow\quad \Sigma,\,\Psi_2 \sim O(\epsilon),\label{smallfluc}
\end{align}   
holds for a given evolution range, with the term $O(\epsilon)\ll 1$ denoting quantities of the order of magnitude of $\epsilon$. Notice that the covariant objects $\Sigma,\,\Psi_2$ vanish at the FLRW background, hence the fluctuations in (\ref{smallfluc}) also vanish at the background, and thus (from Stewart lemma \cite{stewart,ellisbruni89}) are gauge invariant quantities. We derive below the parameter restrictions that yield the necessary and sufficient conditions for a linear regime defined by (\ref{smallfluc}). 

The necessary (not sufficient: see \eqref{linszek}) condition for a Szekeres linear regime is the existence of a linear regime in the LTB seed model, which requires (see \cite{perts}) the radial initial conditions \eqref{initconds1} satisfying for $A=\rho,\,\Theta,\,\KK$
\begin{align}
 \left |A_{qi}-\bar A_i \right |&\sim O(\epsilon),\qquad  \left |rA'_{qi}\right |\sim O(\epsilon)\notag \\
 &\Rightarrow\quad  \drho_i,\,\ddKK_i,\,\ddTh_i\sim O(\epsilon),\label{linreg1}
\end{align} 
where we used the fact that the initial fluctuations of the LTB seed model are linked by the constraint (\ref{constraints2}) restricted to $\bW=0$ and evaluated at $t=t_i$.  As a consequence of (\ref{linreg1}) and bearing in mind \eqref{DaDrho1}--\eqref{qperts}, (\ref{pertseq1})--(\ref{pertseq2}) and the results of \ref{AppB}, we have up to $O(\epsilon)$
\begin{align}
\frac{A_q}{A},\,\, &\frac{A_q}{\bar A},\,\, \frac{A}{\bar A},\,\, \frac{a}{\bar a},\,\, \Gamma \approx 1 \notag\\ 
 \, \Rightarrow\, &\Drho\approx \Drhoas,\,\, \DDTh\approx \DDThas,\,\,\DDKK\approx \DDKKas,\label{flucequiv}
\end{align} 
which suggests introducing the following notation valid up to $O(\epsilon)$: 
\begin{align}
\Drholin=\Drho=&\Drhoas,\,\,\DDalin=\DDa=\DDaas,\notag \\ 
&a_1 \equiv a-\bar a,
\end{align}
that will be used henceforth to denote both types of fluctuations  we have used so far, as they are indistinguishably when linearised.

The linearised forms for the metric functions (see derivation in \ref{AppB}\footnote{The derivation of all functions appearing in the analytic forms of $\Gamma-1$ and $a_1=a-\bar a$ and their linear expansions are given in \ref{AppA} and \ref{AppB}. Equations \eqref{DDrholin}--\eqref{DDThlin} simplify considerably if the decaying mode is suppressed (which imposes a link between $\drho_i$ and $\ddkk_i$).}) are
\begin{align}
a_1=&a-\bar a \approx \bar\phi^m(\hat\Omega_{qi}^m-\bar\Omega_i^m)+\bar\phi^k(\hat\Omega_{qi}^k-\bar\Omega_i^k)\sim O(\epsilon),\label{aLin}\\
&\Gamma-1 \approx -\bar\phi^m\drho_i-\bar\phi^k\ddkk_i\sim O(\epsilon),\label{GLin}
\end{align}
where $\bar\phi^m(\bar a),\,\bar\phi^k(\bar a)$ are dimensionless functions of $O(1)$ defined in (\ref{barPhi})
\begin{equation}\hat\Omega^m_{qi}(r)=\frac{8\pi\rho_{qi}}{3\bar H_i^2}=\bar\Omega_i^m\frac{\rho_{qi}}{\bar\rho_i},\qquad \hat\Omega^k_{qi}(r)=\frac{\KK_{qi}}{\bar H_i^2}=\bar\Omega_i^k\frac{\KK_{qi}}{\bar\KK_i},\label{hatOms}\end{equation}
where $\bar\Omega_i^m=8\pi\bar\rho_i/(3\bar H_i^2$ and $\bar\Omega_i^k=\bar\KK_i/\bar H_i^2=\bar\Omega_i^m+\bar\Omega_i^\Lambda-1$ are the standard density fraction FLRW parameters. 
\begin{widetext}
To obtain the sufficient condition we expand (\ref{DDrho})--(\ref{DDTh}) in terms of $\drho_i,\,\ddKK_i$ which (from (\ref{linreg1})) are $\sim O(\epsilon)$ quantities. Considering that $\hat\Omega_{qi}^m-\bar\Omega_i^m$ and $\hat\Omega_{qi}^k-\bar\Omega_i^k$ are both $\sim O(\epsilon)$, together with (\ref{aLin})--(\ref{GLin}), we obtain:
\ba 
\DDrho_1 &\approx & \frac{\bar\rho_i}{\bar a^3}\,\frac{(1+\bar\phi^m)\drho_i+\bar\phi^k\,\ddkk_i}{1-\bW}\left[1+ O\left(\frac{\epsilon}{1-\bW}\right)\right],\label{DDrholin}\\
 \DDTh_1 &\approx & \frac{\frac32\bar\Omega_i^m[(1+\bar\phi^m)\drho_i+\bar\phi^k\,\ddkk_i]-\bar a\left[\bar\Omega_i^k(\bar\phi^m\drho_i+\bar\phi^k\ddkk_i)+\frac32\ddkk_i\right]}{(\bar\Theta/3)\,\bar a^3\,(1-\bW)\,\bar H_i^{-1}} \left[1+ O\left(\frac{\epsilon}{1-\bW}\right)\right], \label{DDThlin}
\ea          
where (and this is important to notice) we did not restrict the dipole term $\bW$ to be small ({\it i.e}, we have in general $|\bW|\sim O(1)$).

For purely growing modes (see \ref{AppA2}), the suppression of the decaying mode yields a constraint between the initial fluctuations $\drho_i$ and $\ddkk_i$. Therefore,  the expansions \eqref{DDrholin} and \eqref{DDThlin} take the simplified form
\ba 
\DDrho_1 &\approx & \frac{\bar\rho_i}{\bar a^3}\, \frac{(1+\bar\FF)\,\drho_i}{1-W}\left[1+ O\left(\frac{\epsilon}{1-\bW}\right)\right],\label{DDrholin2}\\
\DDTh_1 &\approx & \frac{\left[\frac{3}{2}\bar\Omega_i^m(1+\bar\FF)-\bar a\left(\bar\Omega_i^k\bar\FF-\frac{3}{2}\bar\Phi_i^m/\bar\Phi_i^k\right)\right]\drho_i}{\bar a^3(\bar\Theta/3)(1-\bW)\bar H_i^{-1}}\left[1+ O\left(\frac{\epsilon}{1-\bW}\right)\right],\label{DDThlin2}
\ea
where the background quantities $\bar\FF(\bar a),\,\bar\Phi_i^m,\,\bar\Phi_i^k$ are defined in \ref{AppB}. Their explicit forms for a $\Lambda$CDM background ($\bar\Omega_i^k=0$) are given by \eqref{PhimLCDM}--\eqref{Gminus1}.    
\end{widetext}
The necessary and sufficient condition for a linear regime in generic Szekeres models are then the necessary conditions (\ref{linreg1})--(\ref{GLin})  for the linear regime of the seed LTB model plus the extra condition involving the dipole term:
\ba \frac{\epsilon}{1-\bW}\ll 1\qquad \Rightarrow\qquad \epsilon \ll 1-\bW.\label{linszek}\ea
It is important to emphasise that a linear regime in Szekeres models, as specified by (\ref{linreg1}) and (\ref{linszek}), does not imply closeness to spherical symmetry ({\it i.e.} an ``almost spherical model'' complying with $|\bW|\ll 1$). As long as (\ref{linszek}) holds, large local deviations from spherical symmetry associated with small $1-\bW$ (see \cite{multi,coarse}) are perfectly compatible with a linear regime.

\section{Linearised evolution equations and their solutions.}\label{Lineveqs}

Applying the criterion for a linear regime given by (\ref{smallfluc}) to the system (\ref{as1})--(\ref{c2ndorder}) we obtain its linearised form consisting of:
\begin{itemize} 
\item FLRW background equations: are identical to (\ref{as1}), (\ref{as2}) and (\ref{c1as})
\begin{align}
\dot{\bar \rho} = &-\bar\rho\,\bar \Theta,\,\,\notag\\
\dot{\bar\Theta} =& -\frac{\bar\Theta^2}{3}-4\pi\bar\rho
+8\pi\Lambda,\label{linbackgr}\\
\left(\frac{\bar\Theta}{3}\right)^2 =&\frac{8\pi}{3}\left[\,\bar\rho+\Lambda\,\right]-\bar\KK,\notag 
\end{align}
\item Linearised evolution equations for the fluctuations $\Drhoas,\,\DDThas$ (linearised forms of (\ref{as3})--(\ref{as4}))
\ba \dot\Drholin &=& -\DDThlin,\label{lin2a}\\
\dot {\textrm{\bf{D}}}^{(\Theta)}_1 &=& -\frac{2}{3}\bar\Theta\,\DDThlin-4\pi\bar\rho\,\Drholin,\label{lin2b}
\ea
\item Constraint that defines the spatial curvature fluctuation (notice that in (\ref{c2as}) we have $(\Theta_q-\bar\Theta)^2\sim O(\epsilon^2)$)
\ba\frac{3}{2}\DDKKlin = {4\pi}\bar\rho\Drholin-\frac{\bar\Theta}{3}\DDThlin.\label{linc2as}\ea
\item Linearised form of the second order equation (\ref{c2ndorder}) for the density contrast fluctuation
\ba\ddot\Drholin+\frac{2}{3}\bar\Theta\,\dot\Drholin-4\pi\bar\rho\,\Drholin=0,\label{lin2ndorder}\ea 
\end{itemize}
In what follows we examine the analytic solutions for this linearised system by assuming initial conditions at the last scattering surface ($t_i=\tls,\,\,z\sim 1100$), so that the $\Lambda$CDM background is very close to an Einstein--de Sitter background model ($\KK_i=\Lambda=0$, see \ref{AppB}). The solutions follow  by applying the approximations (\ref{apprlin}) to the exact forms (\ref{Drhoexact})--(\ref{Cdexact}) (see \ref{AppB}) and by bearing in mind together the equivalences (\ref{flucequiv}):
\ba \Drholin &=& \Cg\,\bar a+\frac{\Cd}{\bar a^{3/2}},\label{linsols1}\\
\hbox{with} &&\Cg=-\frac{3}{5}\frac{\DDKKLS}{\Hls^2},\qquad \Cd=\DrhoLS+\frac{3}{5}\frac{\DDKKLS}{\Hls^2},\notag
\ea 
and where the subindex ${}_{1\textrm{\tiny{LS}}}$ denotes $O(\epsilon)$ quantities evaluated at $t=t_i=\tls$ and $\bar a^{3/2}=(3/2)\Hls(t-\tls)+1$ (so that $t=\tls$ corresponds to $\bar a =\bar a_i=1$). The coefficients $C_{\pm}=C_{\pm}(r,\theta,\phi)$ identify the amplitudes of the growing ($+$) and decaying ($-$) modes. The remaining fluctuations follow from (\ref{lin2a}) and (\ref{linc2as})
\ba \frac{\DDThlin}{\Hls}=-\frac{\Cg}{\bar a^{1/2}}+\frac{3}{2}\frac{\Cd}{\bar a^3},\qquad \frac{\DDKKlin}{\Hls^2}=\frac{5}{3}\frac{\Cg}{\bar a^2}.\label{linsols2}\ea
where we remark that (as expected) the linearised curvature fluctuation has no contribution from the decaying mode. 

It is customary to eliminate the decaying mode by setting $\Cd=0$
\footnote{As argued in \cite{multi,coarse}, it is not strictly necessary to totally eliminate the decaying mode, which is equivalent to demanding a perfectly simultaeous Big Bang ($\tbb'=0$). Linear initial conditions imply that a small amplitude decaying mode (of the order of initial fluctuations) produces a gradient $r\tbb'$ that leads to age differences ($\sim 10^3-10^4$ years) among observers that are negligible in comparison with cosmic age.}
, which yields the following constraint linking the density and curvature fluctuations 
\ba 
\label{growing:mode}
\DDKKLS=\frac{5}{3}\Hls^2\DrhoLS,
\ea 
and thus the pure growing mode solutions of (\ref{lin2ndorder}) are (\ref{linsols1})--(\ref{linsols2}) with $\Cd=0$:
\begin{equation}
\Drholin = \DrhoLS\,\bar a,\,\, \DDThlin=-\frac{\DrhoLS}{\bar a^{1/2}}\,\Hls,\,\,\DDKKLS=\frac{5}{3}\frac{\DrhoLS}{\bar a^2}\,\Hls^2,\label{linsols3}
\end{equation}  
with $\DrhoLS=\Cg$. While these linear fluctuations are the Szekeres analogues of linear CPT fluctuations, there are important and subtle differences: they are deterministic while linear CPT perturbations are based on random field variables {which contain the former. This is because the solutions in \eqref{growing:mode} are separable and thus the evolution is independent of the initial configuration. This universal evolution of the linearised fluctuations is known as the transfer function}. We discuss this issue in section 10, in particular we compare $\Drho_1$ with the matter density CPT perturbation $\delta_1$.   


\section{Szekeres models $\&$ Cosmological Perturbation Theory: equivalence of equations.}

The metric in the Cosmological Perturbation Theory formalism in the isochronous gauge can be written in a similar form as the Szekeres metric in \eqref{szmetric} (for more details see e.g. \cite{EMM}):
\begin{equation}
\label{perts:interval}
ds^2=\bar{a}^2(\tau)[-d\tau ^2 +\gamma_{ij}(\mathbf{x},\tau)dx^i dx^j],
\end{equation}
where $\gamma_{ij}$ is the 3-metric or conformal spatial metric and $\tau$ is the conformal time related to physical cosmic time by $\tau=\int{\dd t/\bar a(t)}$. The density contrast $\delta$ is defined by: 
\begin{equation}
\rho(\mathbf{x},\tau)=\rho(\tau)+\delta \rho(\mathbf{x},\eta)=\bar{\rho}(\tau)(1+\delta(\mathbf{x},\tau)),
\end{equation}
with $\bar{\rho}$ denoting the background density. The deformation tensor $\vartheta^\mu_\nu$ is given by:
\begin{equation}
\vartheta^\mu_\nu= \bar{a} \, u^\mu_{\;;\nu}-\frac{\bar{\Theta}}{3}h^\mu_{\;\nu},\qquad \frac{\bar{\Theta}}{3} = \bar{H}=\frac{1}{\bar{a}}\frac{d\bar{a}}{d\tau},\label{vartheta}
\end{equation}
where the isotropic background expansion is given by the conformal Hubble scalar $\bar{H}(\tau)$ and the projection tensor $h^\mu_{~\nu}  = u^\mu u_\nu + \delta^\mu_{~\nu}$ must be computed with $u^\mu=\bar a(\tau)\delta^\mu_\tau$. 

The evolution equations for the variables $\delta$ and $\vartheta^\mu_\nu$ are furnished by the continuity and Raychaudhuri equations: 
\begin{eqnarray}
\frac{\partial\delta}{\partial\tau}+(1+\delta)\vartheta=0,\label{ContEqPert}
\\
\frac{\partial\vartheta}{\partial\tau}+\frac{\bar{\Theta}}{3}\vartheta +\vartheta^{\mu}_{\nu}\vartheta^{\nu}_{\mu}+4\pi G\bar{a}^2\bar{\rho }\delta =0,\label{RayEqPert}
\end{eqnarray}
with $\vartheta=\vartheta^{\mu}_{\;\mu}$. 
%
Since $\tau=\tau(t)$ and thus $\partial/\partial\tau=\bar a(t)\,\partial/\partial t$, and considering that $u^\mu_{;\nu}=\sigma^\mu_\nu+(\Theta/3)h^\mu_\nu$ holds for Szekeres models (as the 4--acceleration and vorticity associated with $u^\mu$ vanish), we can rewrite the CPT deformation tensor introduced in \eqref{vartheta} and its trace $\vartheta$ in terms of variables we have used to examine Szekeres models: the shear tensor and the contrast Hubble scalar fluctuation as follows 
\ba
\vartheta^\mu_{~\nu} = \bar{a}(t) \left(\frac13 {\textrm{\bf{D}}}^{(\Theta)}_{(\textrm{\tiny{as}})}h^\mu_{~\nu} + \sigma^\mu_{~\nu}\right),\label{vartheta2}
\\
\vartheta(t)= \vartheta^{\mu}_{\,\mu}(t) = \bar{a}(t)\, {\textrm{\bf{D}}}^{(\Theta)}_{(\textrm{\tiny{as}})},\label{vartheta3}
\\
\sigma^\mu_{~\nu} = \frac13  \left(- \DDThas+\Theta_q - \bar{\Theta}   \right)\,\eb^\mu_{~\nu},\label{sigma2}
\ea
where we used ~\eqref{pertseq2} and \eqref{sigE2}. From here onwards we will compute all quantities in terms of cosmic time (hence we remove the ``$(t)$'' label).

If we rewrite the continuity and the Raychaudhuri equations \eqref{ContEqPert}--\eqref{RayEqPert} in terms of the variables we have used, we find that they match exactly with the corresponding equations for asymptotic fluctuations \eqref{as3}-\eqref{as4} when the following non--linear correspondences between the CPT and Szekeres fluctuations hold: 
\begin{equation}
\label{perts-as:equiv}
\vartheta \leftrightarrow \bar{a}\, \DDThas,\qquad \delta \leftrightarrow  \Drhoas   .
\end{equation}
Since the evolution equations for these non--linear variables are mathematically identical, the perturbative equations to all orders should be also identified. 
We proceed to relate the spatial curvature perturbations with the curvature of asymptotic variables by means of the definition $6 \mathcal{K} = \RR$. The 3--Ricci scalar of the spatial metric $h_{ik}$ can be expressed as
\begin{equation}
\label{K:3R}
\RR = 6  \left[\bar{\mathcal{K}} + \DDKKas\right].
\end{equation}

\noindent Substituting this definition and the correspondences of Eq.~\eqref{perts-as:equiv} in the non--linear Hamiltonian constraint of CPT (see e.g. \cite{ellisbruni89}), 
\begin{equation}
\vartheta^2 +\frac{4}{3}\bar{\Theta}\vartheta -\vartheta^{i}_{j}\vartheta^{j}_{i}+ \RR\bar{a}^2  =  16\pi G\bar{a}^2 \bar{\rho}\delta\,, 
\label{HamEqPert}
\end{equation}

\noindent we recover the homogeneous constraint, Eq. \eqref{c1as}, as well as the fluctuations constraint, Eq. \eqref{c2as}, at the non--linear level.

We have thus related our variables describing Szekeres exact solutions to the covariant and gague-invariant set of variables of Cosmological Perturbation Theory. Let us exploit our new relations and the solutions for the Szekeres variables to show that, through the Hamiltonian constraint, the spatial curvature scalar is time-independent. At first order in perturbations, the constraint in Eq.~\eqref{c2as} drops the last term and it can be written as an expression for the 3--Ricci scalar at first order,
\begin{equation}
\label{linear:hamilton}
\RR_1 \bar{a}^2= 6 \bar{a}^2 \DDKKlin= 16 \pi \bar{a}^2 \bar{\rho} \Drholin - \frac43 \Theta \DDThlin\,.
\end{equation}
We note that the three-Ricci curvature scalar for the conformal metric $\gamma_{ij}$ relates to the Ricci scalar above as\footnote{The tree-Ricci scalar for the full spatial metric $\RR$ is conformal to the three-Ricci scalar for $\gamma_{ij}$, denoted throughout the text by $\RR_{\gamma} $ See Eqns.~\eqref{perts:interval} and \eqref{szmetric}.}:
\begin{equation}
\RR_{\gamma} = a^2\,\times\,\RR,\label{RRgamma1}
\end{equation}
and thus, as a consequence, the time derivative of Eq.~\eqref{linear:hamilton} can be employed to show that the $\RR_{\gamma}$ is constant in time order by order. Up to $O(\epsilon)$ we can write
\begin{eqnarray}
\RR_{\gamma} &=&\left(\bar{a} + a_{1} \right)^2 \left[\bar{\RR} + \RR_1\right] +  
O(\epsilon^2)\,,\nonumber\\
&=& \left(\bar{a}^2 + 2\bar{a}a_{1} \right) \left[6\bar{\mathcal{K}} + 6 \DDKKlin\right]
 +  O(\epsilon^2)\,,\nonumber\\
 &=& 6 \bar{a}^2 \bar{\mathcal{K}} + 12\bar{a}a_{1} \bar{\mathcal{K}} + 6 \bar{a}^2\DDKKlin +  O(\epsilon^2).\label{RRgamma2}
\end{eqnarray}
where $a_1=a_1(t,r)$ is defined in \eqref{aLin}. 
At the homogeneous level the form of $\bar{\mathcal{K}}$ in Eq.~\eqref{flrwscals} immediately shows that $\RR_{\gamma}$ is constant in time in the FLRW background. Using the evolution equations \eqref{as1} and \eqref{as2} for the homogeneous quantities, as well as the linearized evolution of the fluctuations (Eqs.~(\ref{lin2a})--(\ref{lin2b})) we find at first order the following important result
\begin{equation}
\frac{d}{dt} \left[\RR_{1\gamma} \right]= 12 
\frac{d}{dt}  \left(\bar{a}a_1 \bar{\mathcal{K}}\right)- 4 \bar{a}^2 \bar{\mathcal{K}}\, \DDThlin\, = 0.
\end{equation}
which is consistent with the result from perturbation theory \cite{Wands:2000,Lyth:2004,Langlois:2005}, because $\DDKKlin$ dictates the amplitude of the growing mode of the linearized density contrast, as shown explicitly in \eqref{linsols1}--\eqref{linsols2}, and also expected from CPT. Having the time dependence of $\DDKKas$  at hand, we can read the amplitude from the power spectrum at any time and evolve the fluctuations in time with the solutions of Eq. \eqref{linsols2} and, subsequently, with the non--linear solution in the non--perturbative regime.

\section{Growth factor and collapse morpholgies}

The observable that accounts for the growth of structure is the growth factor $f$, usually parametrised (in the linear regime) as a function of the matter density fraction. In the language of CPT,
\begin{equation}
f_1 \equiv \frac{d \ln \delta_1}{d \ln \bar{a}}= \Omega_m^\gamma\,,\label{f1gamma}
\end{equation}
where $\gamma$ is known as the growth index that distinguishes between different gravity theories and background models \cite{RSD2}. However, the proper interpretation of observations is subject to considering non--linear effects in collapsing structures. Since Szekeres solutions are an exact non--linear extension of CPT that comprises all orders of approximation,  we examine in this section how the admissible collapse morphologies associated with these solutions can modify the prescriptions for structure growth at non--linear order. 
 
\subsection{Szekeres collapse morphologies}
The geometry of the collapse of a dust source is dictated by the time evolution along the principal space directions associated with the eigenvalues (and their associated scale factors) of the deformation tensor $\vartheta^\mu_\nu$ defined in \eqref{vartheta}.  From \eqref{vartheta2}--\eqref{sigma2} the deformation tensor for Szekeres models can be easily written in terms of the exact asymptotic fluctuations as
\ba
\vartheta^\mu_\nu = \frac{\bar a}{3}\left[\left(-\DDThas+\Theta_q-\bar\Theta\right)\eb^\mu_{~\nu}+\DDThas\,h^\mu_\nu\right].\label{vartheta4}
\ea
We can identify the three the nonzero eigenvalues of $\vartheta^\mu_\nu$ and their associated scale factors:
\begin{align} 
\vartheta_{(1)}=&\vartheta^\mu_\nu\eb^\mu_{(1)}\eb_\nu^{(1)}= \bar a\,\frac{\dot \ell_{(1)}}{\ell_{(1)}}=\bar a\left[\DDTh+\frac13(\Theta_q-\bar\Theta)\right] ,\label{eigv1}\\
\vartheta_{(2)}=&\vartheta_{(3)}=\vartheta^\mu_\nu\eb^\mu_{(2)}\eb_\nu^{(2)} = \vartheta^\mu_\nu\eb^\mu_{(3)}\eb_\nu^{(3)}=\bar a\,\frac{\dot \ell_{(2)}}{\ell_{(2)}}=\bar a\,\frac{\dot \ell_{(3)}}{\ell_{(3)}}\notag \\
=& \frac{\bar a}{3}(\Theta_q-\bar\Theta),\label{eigv2}\\
 \ell_{(1)} =& \frac{\Gamma-\bW}{1-\bW}\,\frac{\bar a}{\bar a},\qquad 
 \ell_{(2)}=\ell_{(3)}=\frac{a}{\bar a},\label{ells}
 \end{align}
where we used \eqref{FFq6}, and the triad vectors $\eb_{(i)}^\mu$ satisfy $h_{\mu\nu}\eb_{(i)}^\mu\eb_{(j)}^\nu=\delta_{(i)(j)}$.

 Szekeres models admit spherical and pancake types of collapse morphology \cite{barrow-silk,multi,coarse,Meures:2011ke}. By looking at the eigenvalues of $\vartheta^\mu_\nu$ in \eqref{eigv1}--\eqref{eigv2} we can see that a ``pancake'' collapse occurs in an expanding model ($\Theta_q,\,\bar\Theta>0$) along the principal direction marked by $\vartheta_{(1)}$ (see numerical example in \cite{coarse}). This is straightforward to verify: since in regions where $\Gamma-\bW\approx 0$ the scale factor $\ell_{(1)}$ decreases while  $\ell_{(2)}=\ell_{(3)}$ keep growing and the eigenvalue $\vartheta_{(1)}$ also increases,  diverging as a shell crossing is approached  $\Gamma-\bW\to 0$, all this happening as the remaining two eigenvalues $\vartheta_{(2)}=\vartheta_{(3)}$ remain positive and bounded. As a contrast, spherical collapse occurs for regions around $r=0$ with $\Theta_q<0$ where dust layers collapse in an expanding background (thus $\bar\Theta>0$ remains finite). The three eigenvalues \eqref{eigv1}--\eqref{eigv2} and diverge at the Big Crunch collapse singularity when $\Theta_q\to-\infty$ and $\DDTh\to-\infty$ as the three scale factors \eqref{ells} tend to zero (see \eqref{qscals} and \eqref{DDTh}). 

\subsection{The Szekeres growth factor}

The growth factor in linear CTP is defined by the following well known expression valid up to $O(\epsilon)$
\begin{align}
 f_1 =& \frac{d\ln \delta_1}{d \ln \bar{a}} =\frac{\dot\delta_1}{\delta_1\,\bar H},\label{flin}\\
 \bar H =& \frac{\bar\Theta}{3}=\frac{\dot{\bar a}}{\bar a}=\bar H_i\frac{\left[\bar\Omega_i^m-\bar\Omega_i^k\,\bar a+\bar\Omega_i^\Lambda\,\bar a^3\right]^{1/2}}{\bar a^{3/2}},\end{align}
where for a $\Lambda$CDM background we have $\Omega_i^k=0$, hence $\Omega_i^\Lambda=1-\Omega_i^m$. 
Considering the equivalence relations \eqref{perts-as:equiv} between the linear CPT $\delta_1,\,\vartheta_1$ and the exact Szekeres fluctuations $\Drhoas,\,\DDThas$, the exact generalisation of the linear growth factor \eqref{flin} is
\begin{equation} f= \frac{\dot\Drhoas}{\Drhoas\,\bar H}=-\frac{\left[1+\Drhoas\right]\,\DDThas}{\Drhoas\,\bar H},\label{fexact}\end{equation}
where we used \eqref{as3}, \eqref{pertseq1}--\eqref{pertseq2} and \eqref{DDrho}--\eqref{DDTh}. Since the equivalence relations \eqref{perts-as:equiv} are valid at all orders, we examine the growth factor $f$ for Szekeres models at various approximations and compare with its form in CPT.
\begin{figure}[h!]
      \centering
      \includegraphics[width=0.5\textwidth]{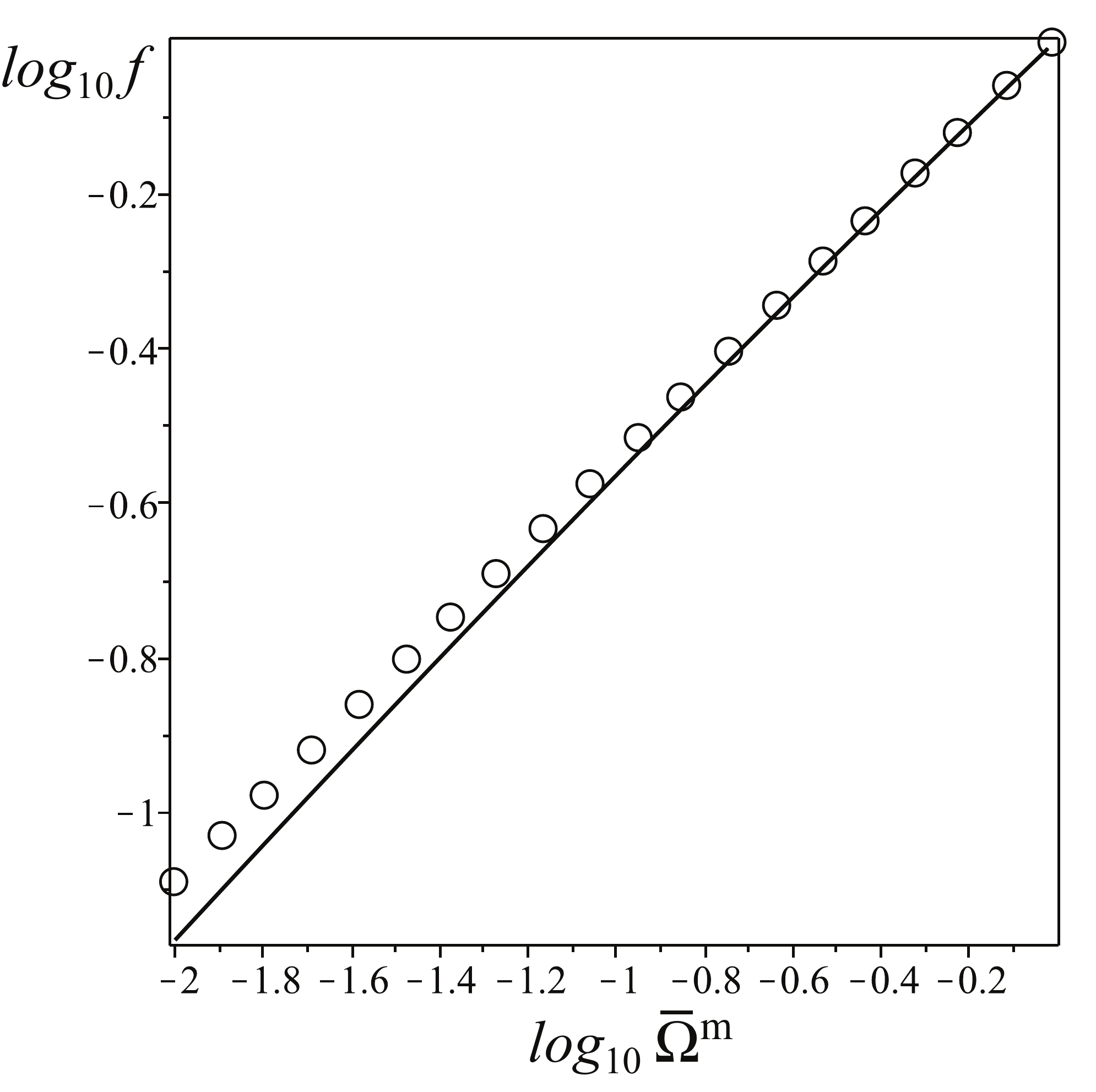}
      \hspace{0.5cm}
       \caption{ The linear suppression factor. Logarithmic graph of the Szekeres linear growth factor $f_1$ obtained in \eqref{flin2} vs the $\Lambda$CDM background parameter $\bar\Omega^m$, both expressed as functions of $(\bar a,\bar\Omega_0^m)$ for the values $\Omega_0^m=0.2,\,0.35$. The line with slope $6/11$ is represented by the circles. The figure shows how $f_1\sim [\bar\Omega^m]^{6/11}$ is a good approximation for $f_1$ in the ranges $\bar\Omega^m\approx 1$. }\label{fvsOmega}
\end{figure}    

\begin{description}
\item[Linear regime] At first order the growth factor \eqref{fexact} takes the form 
\begin{equation} f_1= -\frac{\DDTh_1}{\Drho_1\,\bar H}, \label{flin1}\end{equation}
where $\DDTh_1$ and $\Drho_1$ can be computed from the leading terms in \eqref{DDrholin}--\eqref{DDThlin} (or \eqref{DDrholin2}--\eqref{DDThlin2}) together with the form for $1-\Gamma$ derived in \ref{AppB}. As long as the decaying mode is suppressed (see \ref{AppA2} and \ref{AppB}), for whatever FLRW background that may be chosen the linearised factor $f_1$ exhibits the characteristic features of the linear CPT growth factor: it is necessarily a bounded quantity that is insensitive to collapse morphologies and depends only on background variables, since the initial fluctuations $\drho_i$ and $\ddkk_i$ cancel out (because of \eqref{Gammaexact2} and \eqref{Gminus1}) and the terms $a^2,\,\,a^3$ and $1-\bW$ in the denominators of \eqref{DDrholin}--\eqref{DDThlin} also cancel out. In particular, for a $\Lambda$CDM background with suppressed decaying mode with $t_i=t_0$ (present cosmic time) we have (from \eqref{DDrholin2}--\eqref{DDThlin2})
\ba  f_1= \frac{3}{2} \frac{\bar a- \bar\Omega_0^m\bar\Phi^k(\bar a)}{\left[\bar\Omega_0^m+(1-\bar\Omega_0^m)\bar a^3\right]\,\bar\Phi^k},\label{flin2}\ea
where ${}_0$ denotes evaluation at $t_0$ and we used the form for $1-\Gamma$ derived in \eqref{Gminus1} with $\bar\Phi^k$ and $\bar\Phi_0^k$ defined in \eqref{PhimLCDM}. As expected (see figure \ref{fvsOmega}), the Szekeres linear growth factor \eqref{flin2} also complies with the well known linear CPT approximation
\begin{equation} f_1\approx[\bar\Omega^m]^{6/11},\label{f1aprox}\end{equation} 
where $\bar\Omega^m(\bar a)=\bar\Omega_0^m/[\bar\Omega_0^m+(1-\bar\Omega_0^m)\bar a^3]$. This result has been taken as a probe of gravity, arguing that deviations from the growth index $\gamma = 6/11$ imply a departure from the GR prescription \cite{Lahav-Rees,RSD1,Bueno-Bellido}. 
\item[Non--linear approximation and exact form] Up to second order \eqref{fexact} becomes 
\begin{equation} f_2 = -\frac{(1+\Drho_1)\DDThlin+\DDTh_2}{\Drho_1\bar H},\label{f2}\end{equation}
where the second order term $\DDTh_2$ can be computed from \eqref{DDThlin} or \eqref{DDThlin2}. It is evident that the non--linear (but still perturbative) $f_2$ is now sensitive to the collapse morphologies, since \eqref{f2} is no longer a simple quotient of fluctuations, and thus, even if suppressing the decaying mode, the initial fluctuations and terms $a^2,\,\,a^3$ and $1-\bW$ in the denominators of \eqref{DDrholin}--\eqref{DDThlin} no longer cancel out (The morphology-dependence in the growth function is also manifest in second--order CPT \cite{Brunietal}). This sensitivity to collapse morphologies is even more manifestly evident if we compute the exact form of $f$ in \eqref{fexact}
\begin{widetext}
\begin{equation}
 f = \frac{(\hat\Omega_{q0}^m/\bar\Omega_0^m)(1+\drho_0-\bW)\,\bar a^3}
{(\Gamma-\bW)\,a^3-(\hat\Omega_{q0}^m/\bar\Omega_0^m)(1+\drho_0-\bW)\,\bar a^3} \left[\frac{\frac{3}{2}\hat\Omega_{q0}^m(1-\Gamma+\drho_i)-a\left[\hat\Omega_{q0}^k(1-\Gamma)+\frac{3}{2}\ddkk_0\right]}{(a/\bar a)^{3/2}(\Gamma-\bW)\sqrt{\hat\Omega_{q0}^m-\hat\Omega_{q0}^k a+\bar\Omega_0^\Lambda a^3}\sqrt{\bar\Omega_0^m-\bar\Omega_0^k \bar a+\bar\Omega_0^\Lambda \bar a^3}}-3\left(\frac{\Theta_q}{\bar\Theta}-1\right)\right], 
\label{fexactfull}
\end{equation}
\end{widetext}
where we used the exact forms \eqref{DDrho}--\eqref{DDTh} and their relation with asymptotic fluctuations in \eqref{pertseq1}--\eqref{pertseq2}.
We examine the sensitivity to collapse morphologies by comparing \eqref{fexactfull} and the scale factors \eqref{ells} associated with them. Assuming an expanding (but otherwise generic) FLRW background (so that $\bar a$ is ever growing and $\bar H>0$) we have  
\begin{itemize}
\item Pancake collapse. It is evident that $f$ can exhibit very large growth if $\Gamma-\bW$ becomes sufficiently small for large $a$, hence the scale factor $\ell_{(1)}$ decreases for increasing $\ell_{(2)}=\ell_{(3)}$ and $f$ diverges (shell crossing) as $\ell_{(1)}\to 0$. Likewise, the eigenvalue $\vartheta_{(1)}$ diverges for finite $\vartheta_{(2)}=\vartheta_{(3)}$ as $\Gamma-\bW\to 0$.
\item  Spherical collapse. The growth factor can also increase as $0\approx a\ll 1$ for bounded $\Gamma-\bW$, so that the three scale factors $\ell_{(1)}, \ell_{(2)}=\ell_{(3)}$ decrease while the eigenvalues   $\vartheta_{(1)}, \vartheta_{(2)}=\vartheta_{(3)}$ diverge, a collapse singularity occurs as $a\to 0$ (all this happening with increasing and large $\bar a$ and $\bar H$).  
\end{itemize}
Evidently, the behaviour of the exact growth factor \eqref{fexactfull} should be compared with the non--linear second order form \eqref{f2}. In particular, it is necessary to compare both forms \eqref{f2} and \eqref{fexactfull} with the growth factor occurring in previous work using relativistic non--linear perturbations \cite{CPT22,CPT23,CPT24}) and its comparison with the second order Newtonian solution worked out for a $\Lambda$CDM background in the synchronous and comoving gauge \cite{Brunietal}. 
 
\end{description}
          
\section{Connection to metric based perturbations.}

The conditions for a linear regime (\ref{linreg1})--(\ref{GLin}) and (\ref{linszek}) allow us to express the Szekeres line element (\ref{szmetric}) as the metric of a Szekeres model that is close (up to $\sim O(\epsilon)$) to an FLRW model. Considering  that the strict FLRW limit of the models follows from demanding that $\sigma^\mu_\nu=E^\mu_\nu=0$ holds everywhere, the conditions for this limit are (from (\ref{sigE2})--(\ref{LTBperts})) given by 
\begin{align}
\KK'_{qi}&=\rho'_{qi}=0\quad\Rightarrow\quad a'=0,\,\,\Gamma=1\nonumber\\
 &\Rightarrow \quad \DDrho=\DDKK=\DDTh=0,\quad a=\bar a(t).\label{SzFLRW}
\end{align} 
Applying these conditions to (\ref{szmetric}) necessarily transforms this metric, for whatever choice of dipole parameters in $\bW\ne 0$, into a FLRW metric in an unusual coordinate representation. This FLRW metric is
\begin{equation}
\dd s^2 =-\dd t^2+\bar h_{ij}\,\dd x^i\,\dd x^j,\, \bar h_{ij}=\bar a^2\,\bar\gamma_{ij}\, \label{fullszFmetric}
\end{equation}
with $\bar \gamma_{ij}$ given by\footnote{It is straightforward to verify with a computer algebra system that the 4--dimensional Weyl tensor vanishes identically for the metric (\ref{fullszFmetric})--(\ref{spszFmetric}).}
\ba   
\bar \gamma_{ij}\,\dd x^i\,\dd x^j = &&\left[\frac{(1-\bW)^2}{1-\bar\KK_{i}r^2}+(\PP+\bW_{,\theta})^2+U^2\PP_{,\phi}^2\right]\dd r^2-\notag \\ 
&&2r\,(\PP+\bW_{,\theta})\dd r\dd\theta - \nonumber\\
  &&2r\,U\,\PP_{,\phi}\dd r\dd\phi + r^2\left(\dd\theta^2+\sin^2\theta\,\dd\phi^2\right).\label{spszFmetric}
  \ea
Comparing (\ref{spszFmetric}) and (\ref{szmetric1})--(\ref{szmetric3}), we notice that the components of $\bar \gamma_{ij}$ coincide with those of $\gamma_{ij}$ save for $\bar \gamma_{rr}$ and $\gamma_{rr}$. Hence, we can rewrite the full exact Szekeres metric as an FLRW metric plus a ``correction''
\begin{align} 
\dd s^2 &= -\dd t^2+h_{ij}\dd x^i\dd x^j,\nonumber\\
 \hbox{with}&\,\, h_{ij}= a^2\,\left\{\bar \gamma_{ij}+ \left[\frac{(\Gamma-\bW)^2}{1-\KK_{qi}r^2}-\frac{\bar a^2}{a^2}\frac{(1-\bW)^2}{1-\bar\KK_{i}r^2}\right]\delta_i^r\delta_j^r\right\}.
 \end{align}
This way of expressing the Szekeres metric is very useful to obtain its linearised version by applying the conditions for a linear regime. Assuming a $\Lambda$CDM background (hence $\bar\KK_i=0$), together with  $|\KK_{qi}|/\bar H_i^2=|\hat\Omega_{qi}^k|\sim O(\epsilon)$, as well as (\ref{aLin})--(\ref{GLin}) and (\ref{linszek}), leads to   
\begin{align}
\dd s^2=&-\dd t^2+ \bar a^2\,\left[\bar\gamma_{ij}+ G\,\delta_i^r\delta_j^r\right]\dd x^i\dd x^j,\label{szmetriclin}\\
 \hbox{with}&\,\, G(t,x^k) = (1-\bW)\left[2(\Gamma-1)+(1-\bW)\KK_{qi}r^2\right],\label{Glindef}
\end{align}
which must be compared with the generic linearly perturbed metric for dust sources in the isochronous gauge \cite{EMM,CPT22,CPT24,Brunietal}: 
\ba   \dd s^2=-\dd t^2+ \bar a^2\,\left[(1-2\psi)\delta_{kl}+\chi_{kl} \right]\dd y^{k}\dd y^{l}, \label{cptliny}
\ea 
where $\chi_{kl}=\left(\partial_k\partial_l-\frac{1}{3}\nabla^2\delta_{kl}\right)\chi,$ and where the metric potentials $\psi$ and $\chi$ depend on $t$ and on the spatial coordinates $y^{k}$, which are (in general) distinct from $x^i$. In order to obtain these  metric potentials in terms of the Szekeres metric functions $\bar\gamma_{ij}$ and $G$, it is useful to rewrite $\bar\gamma_{ij}$ in (\ref{spszFmetric}) (with $\bar\KK_i=0$) as follows
\ba \bar\gamma_{ij}\dd x^i\dd x^j= \left[\bar\gamma_{ij}^{(0)}+\bar\gamma_{ij}^{(1)}\right]\dd x^i\dd x^j,\label{bargamma}\ea
where
\begin{align}
\bar\gamma_{ij}^{(0)}\dd x^i\dd x^j&=&\dd r^2+r^2(\dd\theta^2+\sin^2\theta\dd\phi^2),\label{bargamma0}\\
\bar\gamma_{ij}^{(1)}\dd x^i\dd x^j &=& (\bar\gamma_{rr}-1)\dd r^2+2\bar\gamma_{r\theta}\dd r\dd\theta+2\bar\gamma_{r\phi}\dd r\dd\phi.\label{bargamma1}
\end{align}
Since $\bar\gamma_{ij}^{(0)}$ is the 3--dimensional flat space metric in spherical coordinates $x^i=r,\theta,\phi$, it becomes a Kronecker delta under the transformation $r^2=x^2+y^2+z^2,\,\,\theta=\arccos(z/r),\,\,\phi=\arctan(y/x)$ into cartesian coordinates $y^k=x,y,z$. Inserting (\ref{bargamma})--(\ref{bargamma1}) into (\ref{szmetriclin}) and transforming into these cartesian coordinates we obtain, after some algebraic manipulation, the linearised Szekeres metric in the form (\ref{cptliny}) with:
\ba 
&&\psi = -\frac{1}{6}\left(\bar\gamma^{(1)}_{rr}+G\right),\label{psi}\\
 &&\left(\partial_k\partial_l-\frac{1}{3}\nabla^2\delta_{kl}\right)\chi =\left(\bar\gamma^{(1)}_{rr}+G\right)\left(r_{,k}r_{,l}-\frac{1}{3}\delta_{kl}\right) + \notag \\
 && \quad 2r_{,k}\left(\bar\gamma_{r\theta}\theta_{,l}+\bar\gamma_{r\phi}\phi_{,l}\right),\label{chikl}\ea
where $\delta^{kl}\chi_{kl}=0$ holds, since we have $\delta^{kl}\delta_{kl}=3,\,\,\delta^{kl}r_{,k}r_{,l}=1$ and $\delta^{kl}r_{,k}\theta_{,l}=\delta^{kl}r_{,k}\phi_{,l}=0$ in the right hand side of (\ref{chikl}). The linear differential equations in (\ref{chikl}) yield $\chi(t,y^k)$, while $\psi$ is then found from (\ref{psi}). For the spherically symmetric LTB models we have $\bar\gamma^{(1)}_{ij}=0$ and the metric potentials do not depend on $(\theta,\phi)$, hence (\ref{psi})--(\ref{chikl}) reduce to
\ba  \psi=-\frac{1}{6}G,\quad G = 2(\Gamma-1)+\KK_{qi}r^2,\notag\\
 \left(\partial_k\partial_l-\frac{1}{3}\nabla^2\delta_{kl}\right)\chi=\left(r_{,k}r_{,l}-\frac{1}{3}\delta_{kl}\right)\,G
\ea
which coincides with the result obtained for LTB models in \cite{perts}.

It is straightforward to compute the linearised forms of the kinematic scalar $\vartheta$ and the shear tensor: 
\ba   
\vartheta &=&-6\bar a\dot \psi =\bar a\dot G =2\bar a (1-\bW)\dot\Gamma,\\
  \sigma_{kl} &=&-\bar a\,\dot\chi_{kl}=-\bar a\,\left(r_{,k}r_{,l}-\frac{1}{3}\delta_{kl}\right)\dot G\notag \\
  &=&-2\bar a\,\left(r_{,k}r_{,l}-\frac{1}{3}\delta_{kl}\right)(1-\bW)\dot\Gamma,\ea
which are consistent (at first order) with the exact forms in (\ref{sigE1}), (\ref{DDTh}) and \eqref{vartheta2}--\eqref{vartheta4}, since from (\ref{FFq6}) we have $\DDTh=\dot\GG/\GG=\dot\Gamma/(\Gamma-\bW)\approx \dot\Gamma/(1-\bW)$. Also, the curvature perturbation at first order in \eqref{RRgamma1}--\eqref{RRgamma2} 
\begin{align}
 \RR{}_{\gamma 1}=&4\nabla^2{\cal{R}}_c,\\
 {\cal{R}}_c=&\psi+\frac{1}{6}\nabla^2\chi=-\frac{1}{6}\left(\bar\gamma^{(1)}_{rr}+G\right)+\frac{1}{6}\nabla^2\chi, \notag
\end{align}
is expressible in terms of the metric potentials once (\ref{chikl}) is solved for a given Szekeres model.

\section{Discussion and conclusion}

We have examined in full detail and rigour the relation between  cosmological perturbation theory (CPT) and the Szekeres models, which (as argued in the Introduction) are the less idealised class of exact solutions of Einstein's equation for the description of cosmic structures. We summarise below the key points and results of this article.

\begin{description}
\item[Szekeres models in coordinate independent variables.]  We have described the dynamics of the models in terms of the coordinate independent variables derived in \cite{sussbol} and used in \cite{multi,coarse} (see sections 2--4). These variables, which provide a natural framework to compare with CPT (the traditional metric based variables used in previous work on the models are not suitable for this purpose), are 
\begin{itemize}
\item  Weighted average functions $A_q$ of the standard covariant scalars $A=\rho,\,\Theta,\,\KK$ (density, Hubble scalar and spatial curvature), 
\item FLRW background variables $\bar A$ that emerge as asymptotic limits of $A_q$ ({\it i.e} when the averaging domain covers the whole time slice), 
\item Fluctuations of the scalars $A$ with respect to the averages $A_q$ ($\Drho,\,\DDa$) and with respect to $\bar A$ ($\Drhoas,\,\DDaas$). In particular, the latter fluctuations (which we denote as ``asymptotic'') naturally provide the exact Szekeres analogue of the density contrast and of the Hubble and curvature perturbations of CPT. However, the fluctuations $\Drho,\,\DDa$ are easier to compute and thus have been kept as useful auxilliar variables. Since both types of fluctuation are equivalent up to $O(\epsilon)$, we denoted them generically by $\Drho_1,\,\DDa_1$ when used in the linear regime (see section 5).
\end{itemize}      
\item[The Szekeres linear regime vs. linear CPT.] We have defined (sections 5--6) rigorously   linear regime for the Szekeres models in terms of a coordinate independent departure from an arbitrary FLRW background and, in particular, from a concordance $\Lambda$CDM cosmology (see \ref{AppB}). We have shown that the linearised Szekeres evolution equations and their solutions fully coincide with the linear CPT evolution equations and their solutions in the isochronous comoving gauge. We have also found (section 9) the relation between the linearised Szekeres metric and the metric potentials of linear CPT for a dust source in this gauge. This equivalence allows us to reproduce and extend results of linear CPT (in any gauge) into the exact Szekeres non--perturbative regime.
\item[Comparison with CPT at all orders.] We proved in section 7 that the correspondence between CPT and the Szekeres models is not only valid in the linear regime (as described above), but is valid at all orders of approximation of CPT in the isochronous comoving gauge.   
\item[Conservation of the curvature perturbation.] Having established the correspondence between Szekeres and CPT variables at all orders, we can prove that well known properties of CPT at a given order also hold for Szekeres models at same order. In particular, we proved in section 7 that the time preservation of the curvature perturbation at all scales in a $\Lambda$CDM background at first order also holds for the linear regime in Szekeres models compatible with this background. 
\item [The Szekeres growth factor.] We derived an exact Szekeres expression that generalises the growth factor used in CPT derived from the study of redshift space distortions (see section 8). We showed how this expression, in the conditions of a Szekeres linear regime and for a $\Lambda$CDM background, reduces to the linear CPT growth factor and thus can also be approximated as as power law form of Eq. \eqref{f1gamma} with exponent $6/11$ (see Eq. \eqref{f1aprox} and figure 1). We also showed how, for the Szekeres non--linear second order approximation to CPT and for the exact non--perturbative regime, the Szekeres growth function becomes very sensitive to the collapse morphologies (``pancake'' and spherical collapses) associated with Szekeres models and discussed in \cite{coarse}. These results suggest that non--linear growth of structures under full GR may produce departures from the linear parametrisation of the growth factor in Eq. \eqref{f1gamma} that should be compared with those prescribed by modified gravity.  
\end{description}
In previous work we have shown \cite{multi,coarse} how Szekeres models describe the evolution of multiple elaborated networks of exact non--linear cosmic structures from early times linear initial fluctuations defined at the last scattering surface (as dust models are not valid for earlier radiation dominated stages). The amplitudes of such fluctuations are fully determined by initial conditions given in terms of Szekeres variables in a linear regime shown explicitly in section 6 (cf.~\eqref{linsols1}--\eqref{linsols3}). On the other hand, the amplitude of cosmological perturbations in CPT, e.g.~the linear matter density perturbation $\delta_1({\bf x},t)$ in the real space, is a Fourier transform of a series of independent harmonic oscillators $\delta_{1{\bf k}}(t)$ with an average amplitude given by the evolved power spectrum as:
\begin{align}
  \delta_1({\bf x},t) =& C({\bf x})\,\bar a(t),\notag \\
  \hbox{with}&\quad C({\bf x})=\frac{1}{(2\pi)^3} \int C_{1{\bf k}} ({\bf x}) \exp(i{\bf k\cdot x})d^3 k,\label{delta1}
\end{align}
which should be compared with the fully determinist form of $\Drho_1$ in \eqref{linsols3}. The independence of modes $C_{1{\bf k}}$ ensures that the real-space amplitude is also a random (Gaussian) variable with an amplitude given by the variance of perturbations. Such random field contains all of the possible Szekeres configurations (among all other possible configurations excluded by the constraints of Szekeres geometry). {This is due to the fact that, in the linear level, the solution of the differential equation for the density contrast is separable and thus independent of the initial amplitude. The time-dependent part of this solution is precisely the transfer function factor of the Powerspectrum.}  While the Szekeres solutions describe only a reduced subset of all possible evolutions, the Szekeres linear density fluctuation \eqref{linsols3} can be evolved (deterministically) into a full exact form, whereas the far more general form \eqref{delta1} is only valid in a linear CPT regime. The fraction of inhomogeneities described by Szekeres solutions is given by the probability of specific real-space configurations, integrated from the probability distribution function (see e.g. \cite{Seery-hidalgo} and  \cite{hidalgo-Polnarev} for an application to configurations collapsing onto primordial black holes). Hence, the correspondence between the linear perturbation from CPT $\delta_1$ in \eqref{delta1} and the linearised Szekeres fluctuation  $\Drho_1$ in \eqref{linsols3} allows us to explore the diversity
\section*{Acknowledgments}

All authors acknowledge financial support from the research grants CONACYT/239639 and CONACYT/269652. We also acknowledge support from grant PAPIIT--UNAM IA-103616. IDG acknowledges financial support from the grants program for postgraduate students of CONACYT.         
\onecolumngrid
\begin{appendix}

\section{Quadratures of the Friedman equation.}\label{AppA}

In practically all previous work on the cosmological applications of Szekeres and LTB models (see comprehensive reviews in \cite{kras1,kras2,BKHC2009,BCK2011,EMM}) the dynamics is determined from the solutions of the following Friedman--like equation (\ref{constraints1}) given in dimensionless form as:

Finally, the correspondence between CPT at all orders and Szekeres models may serve to assess how reliable the parametrisation of the growth function in Eq.~\eqref{f1aprox} is as a test of modified gravity once we include non--linear fluctuations estimated for Szekeres solutions modelling (as in \cite{multi, coarse}) structures in a $\Lambda$CDM background. We will explore this important issue in future articles currently under development. 
   
\ba \frac{H_q^2}{\bar H_i^2}= \frac{\dot a^2}{a^2\bar H_i^2}=
\frac{\hat\Omega^m_{qi}-\hat\Omega^k_{qi}\,a+ \bar\Omega_i^\Lambda\,a^3}{a^3},\qquad H_q=\frac{\Theta_q}{3},\label{Friedeq}
\ea
where the functions $\hat\Omega_{qi}(r),\,\hat\Omega_{qi}^k(r)$ are defined in \eqref{hatOms} and $\bar\Omega_i^\Lambda=8\pi\Lambda/(3\bar H_i^2)$. Since all relevant quantities depend on $a,\,\,\Gamma$ and the initial conditions $\hat\Omega_{qi}(r),\,\hat\Omega_{qi}^k(r),\,\bar\Omega_i^\Lambda$, any further work requires solving \eqref{Friedeq} through the following integral quadrature
\ba 
  \bar H_i(t-\tbb)=F(a,r)=\bar H_i\int_{\xi=0}^{\xi=a}{\frac{\dd\xi}{\xi H_q(\xi)}}=\int_0^a{\frac{\sqrt{\xi}\,\dd\xi}{\left[\hat\Omega^m_{qi}-\hat\Omega^k_{qi}\,\xi+ \bar\Omega_i^\Lambda\,\xi^3\right]^{1/2}}},
\label{Fquadr}\ea 
where $\tbb=\tbb(r)$ is the Big Bang time, which can be eliminated in terms of initial conditions as $\tbb=t_i-F_i/\bar H_i$, with $F_i=F(1,r)$ (since $a=a_i=1$ for $t=t_i$ in (\ref{Fquadr}). 

\subsection{General solutions}

The function $\Gamma$ can be obtained from the formal solution (\ref{Fquadr})
\ba  \Gamma =1 - \Phi^m\,\drho_i-\Phi^k\,\ddkk_i - \HH_q\,r\,\tbb' = 1-\phi^m\,\drho_i-\phi^k\,\ddkk_i,\label{Gammaexact}
\ea
where $\ddkk_i=\ddKK_i/\bar H_i^2$, we have used the fact that $\partial F/\partial a=\bar H_i/(a H_q)$, have eliminated the radial gradient $\tbb'$ above from 
\begin{equation} r\bar H_i \tbb' = -r F_i'=-\Phi^m_i\drho_i-\Phi^k_i\ddkk_i.\label{tbbr}\end{equation}
and the functions $\phi^m,\,\phi^k,\,\Phi^m,\,\Phi^k$ are defined as 
\ba 
  \phi^m = \Phi^m - \Phi_i^m,\qquad \phi^k=\Phi^k-\Phi_i^k,\\
  \Phi^m = 3\hat\Omega_{qi}^m\frac{H_q}{\bar H_i}
 \frac{\partial F}{\partial\hat\Omega_{qi}^m}=-\frac{3}{2}\frac{H_q}{\bar H_i}\int_0^a{\frac{\hat\Omega_{qi}^m\,\dd\xi}{\xi^4(H_q/\bar H)^3}}=-\frac{3}{2}\frac{H_q}{\bar H_i}\int_0^a{\frac{\hat\Omega_{qi}^m\,\sqrt{\xi}\,\dd\xi}{\left[\hat\Omega^m_{qi}-\hat\Omega^k_{qi}\,\xi+ \bar\Omega_i^\Lambda\,\xi^3\right]^{3/2}}},\nonumber\\\label{Phim}\\
  \Phi^k = 3\frac{H_q}{\bar H_i}\frac{\partial F}{\partial\hat\Omega_{qi}^k} =\frac{3}{2}\frac{H_q}{\bar H_i}\int_0^a{\frac{\dd\xi}{\xi^3(H_q/\bar H_i)^3}}=\frac{3}{2}\frac{H_q}{\bar H_i}\int_0^a{\frac{\xi^{3/2}\,\dd\xi}{\left[\hat\Omega^m_{qi}-\hat\Omega^k_{qi}\,\xi+ \bar\Omega_i^\Lambda\,\xi^3\right]^{3/2}}},\nonumber\\
 \label{Phik}
 \ea 
so that $\Phi_i^m,\,\Phi_i^k$ follow by evaluating the integrals in \eqref{Phim}--\eqref{Phik} for the upper limit $a=a_i=1$. Formal analytic forms for $\Drho$ and the remaining fluctuations can be obtained by substitution of \eqref{Gammaexact} in (\ref{qperts}) and \eqref{DDrho}--\eqref{DDTh}.

\subsection{Pure growing mode solutions}\label{AppA2}

A simultaneous Big Bang (associated with a suppressed decaying mode \cite{multi}) implies the following constraint among initial fluctuations and a simplified form for $\Gamma$:
\ba  \tbb'=0\quad\Rightarrow\quad \ddkk_i = -\frac{\Phi^m_i}{\Phi^k_i}\drho_i\nonumber\\
 \qquad \Rightarrow\quad \Gamma = 1-\FF\,\drho_i,\qquad \FF = \Phi^m -\frac{\Phi^k}{\Phi_i^k}\Phi_i^m,
\label{Gammaexact2}\ea
Analytic expressions for $\Drho$ and the remaining fluctuations follow by substitution of \eqref{Gammaexact2} in (\ref{qperts}) and \eqref{DDrho}--\eqref{DDTh}

\subsection{Solutions for $\Lambda=0$}

In the case $\Lambda=\bar\Omega_i^\Lambda=0$ the quadrature (\ref{Fquadr}) is expressible in terms of elementary functions, leading from (\ref{qperts}) to the exact form that generalise the LTB expressions found in \cite{sussmodes}
\ba  \Drho= \frac{\Jg+\Jd}{1-\Jg-\Jd},\qquad \Gamma-\bW=(1+\drho_i-\bW)(1-\Jg-\Jd),\label{Drhoexact}\ea
where 
\ba \Jg=\Cg\left(\Psi-\frac{2}{3}\right),\quad \Jd=\Cd\frac{H_q}{H_{qi}},\ea 
are the exact generalisation of the growing and decaying modes of linear dust perturbations (see comprehensive discussion in \cite{sussmodes}), the modes amplitudes are 
\ba  \Cg=3\frac{\Drho_i-\frac{3}{2}\DKK_i}{1+\Drho_i} = 3\frac{\drho_i-\frac{3}{2}\dKK_i}{1-\bW+\drho_i},\label{Cgexact}\\ 
 \Cd=\frac{r\tbb'}{(1+\Drho_i)(1-\bW)}= 3\frac{(1-\Psi_i)\drho_i-\left(1-\frac{3}{2}\Psi_i\right)\dKK_i}{1-\bW+\drho_i},\label{Cdexact}
\ea
where $\dKK_i=\ddKK_i/\KK_{qi}=\ddkk_i/\kappa_{qi}$ and $\Psi=H_q(t-\tbb)$ is given explicitly by
\ba  \Psi=\frac{e_0\sqrt{2-e_0\alpha_q}}{\alpha_q^{3/2}}\left[\ACal(1-e_0\alpha_q)-\sqrt{\alpha_q}\sqrt{2-e_0\alpha_q}\right],\qquad \alpha_q=\frac{2|\hat\Omega_{qi}^k|}{\hat\Omega_{qi}^m}\,a,\label{Psi}\ea
with $\ACal=\arccos$ for $e_0=1,\,\hat\Omega_{qi}^k=\hat\Omega_{qi}^m-1>0$ (elliptic models) and $\ACal=\hbox{arccosh}$ for $e_0=-1,\,\hat\Omega_{qi}^k<0$ (hyperbolic models). For the case $\hat\Omega_{qi}^k=0$ (parabolic models) we have $\Psi=2/3$ and thus $\Jg=0$ for all choices of $\hat\Omega_{qi}^m$.

\section{The metric functions in the linear regime.}\label{AppB}

Under the linear regime conditions (\ref{linreg1}) the relation between the scale factor $a$ and its equivalent background value $\bar a$ (see discussion in Appendix C of \cite{perts}) can be obtained (up to $O(\epsilon)$) in terms of the relation between the quadrature $F$ in (\ref{Fquadr}) and its background limit. Consider $\PP=[a,\hat\Omega_{qi}^m,\hat\Omega_{qi}^k]$ and $\bar\PP=[\bar a,\bar\Omega_i^m,\bar\Omega_i^k]$ as points in the functional parameter phase space associated with \eqref{Fquadr}. Since $F$ depends smoothly on $\PP$ and contains $\bar F$ as the functional limit $\PP\to\bar\PP$, the following limits define the FLRW background parameters 
\begin{equation} \lim_{\PP\to\bar\PP}\hat\Omega_{qi}^m = \bar\Omega_i^m ,\qquad \lim_{\PP\to\bar\PP}\hat\Omega_{qi}^k = \bar\Omega_i^k,\qquad \lim_{\PP\to\bar\PP} H_{qi} = \bar H_i,\label{Omegas}\end{equation}
which, when substituted into \eqref{Friedeq}--\eqref{Fquadr}, yield $a=\bar a(t)$, leading to   
\ba
 \bar F= F(\bar\PP)=\lim_{\PP\to\bar\PP}F(\PP)=\bar H_i\int_0^{\bar a}{\frac{\sqrt{\xi}\,\dd\xi}{\left[\bar\Omega_i^m-\bar\Omega_i^k\xi+\bar\Omega_i^\Lambda\xi^3\right]^{1/2}}},\qquad \bar F_i=\bar F|_{\bar a=1}.\label{barF}
\ea  
Expanding $F$ around this limit up to first order leads after some algebraic manipulation to equation (\ref{aLin}):
\ba a_1=a-\bar a \approx \bar\Phi^m(\bar a)(\hat\Omega_{qi}^m-\bar\Omega_i^m)+\bar\Phi^k(\bar a)(\hat\Omega_{qi}^k-\bar\Omega_i^k)\sim O(\epsilon),\nonumber\ea
where $\bar\Phi^m(\bar a)$ and $\bar\Phi^m(\bar a)$ are order $O(1)$ quantities defined as the FLRW background limits:
\begin{equation}\bar\Phi^m=\lim_{\PP\to\bar\PP}\Phi^m,\qquad \bar\Phi^k=\lim_{\PP\to\bar\PP}\Phi^k,\label{barPhi}\end{equation}
where $\Phi_m(a)$ and $\Phi_k(a)$ are defined by (\ref{Gammaexact}) and the derivatives of $F$ in $\Phi_m(a),\,\,\Phi_k(a)$ must be evaluated before taking the background limits $\PP\to\bar\PP$ and $H_q\to\bar H(t)$. Performing the same first order expansion on $\Gamma-1$ in (\ref{Gammaexact}) yields equation (\ref{GLin})
\begin{equation} \Gamma - 1 \approx -\bar\phi^m\drho_i-\bar\phi^k\ddKK_i\sim O(\epsilon),\end{equation}
where $\bar\phi^m=\bar\Phi^m-\bar\Phi_i^m,\,\,\bar\phi^k=\bar\Phi^k-\bar\Phi_i^k$ and we used \eqref{tbbr} to eliminate $\tbb'$. 
In particular, for a $\Lambda$CDM background $\bar\Omega_i^k=0,\,\,\bar\Omega_i^\Lambda=1-\bar\Omega_i^m$ we have 
\ba  \bar\Phi_m = -\frac{3}{2}\frac{\bar H}{\bar H_i}\int_0^{\bar a}{\frac{\bar\Omega_i^m\,\sqrt{\xi}\,\dd\xi}{\left[\bar\Omega_i^m+(1-\bar\Omega_i^m)\xi^3\right]^{3/2}}},\qquad \bar\Phi_k = \frac{3}{2}\frac{\bar H}{\bar H_i}\int_0^{\bar a}{\frac{\xi^{3/2}\,\dd\xi}{\left[\bar\Omega_i^m+(1-\bar\Omega_i^m)\xi^3\right]^{3/2}}}.\nonumber\\
\label{PhimLCDM}\ea 
It is straightforward to show that $\bar\Phi_m=-1$ holds identically, while  $\bar\Phi_k(\bar a)$ is expressible in terms of elliptic functions (and hypergeometric functions for some particular values of $\bar\Omega_i^m$). For a suppressed decaying mode ($\tbb'=0$) we obtain from \eqref{Gammaexact2} and \eqref{PhimLCDM}  
\begin{equation} 1-\Gamma  \approx \bar\FF\,\drho_i,\qquad \bar\FF = \frac{\bar\Phi^k}{\bar\Phi_i^k}-1.\label{Gminus1}\end{equation}

The linear forms of the density fluctuation  in section \ref{Lineveqs} follow by expanding (\ref{Gammaexact}) and (\ref{Drhoexact})--(\ref{Psi}) around $\alpha_{qi}=\hat\Omega_{qi}^k/\hat\Omega_{qi}^m=0$ (or equivalently $\hat\Omega_{qi}^m=1$) denoting linear conditions near a spatially flat $\Lambda$CDM background at last scattering time $t=\tls$. Notice that $\OmLls=\bar\Omega_0^\Lambda(\bar H_0/\Hls)^2\sim 10^{-9}$, hence for all practical purposes we have at $t=\tls$ an Einstein de Sitter background that justifies using (\ref{Drhoexact})--(\ref{Psi}). We have for $\hat\Omega_{qi}^m\approx 1$ 
\ba  \Psi \approx \frac{2}{3}-\frac{2}{15}\hat\Omega_{qi}^k\,a,\qquad a\approx \bar a\approx \bar t^{2/3},\qquad \bar t=\frac{3}{2}\Hls(t-\tls)-1,\label{apprlin}\ea
where we bear in mind that $\bar H_i=\Hls,\,\,H_q\approx \bar H,\,\,H_q/H_{qi}\approx 1/\bar a^{3/2}$ and $\hat\Omega_{q\textrm{\tiny{LS}}}^m\approx \Ommls \approx 1$ hold for $t\approx \tls$. The linearised form for the density fluctuation is then the sum of modes
\ba \Drho = \Jg+\Jd,\qquad |\Jg|\ll 1,\quad |\Jd|\ll 1,\label{Drhoqlin}\ea
where the functional form of the modes follows by applying the approximations (\ref{apprlin}) to the exact forms (\ref{Drhoexact})--(\ref{Cdexact}).

\end{appendix}


\end{document}